\documentclass[correspondence, draftcls, onecolumn, 12pt]{IEEEtran}
\usepackage[dvips]{graphicx}
\usepackage[fleqn]{amsmath}
\usepackage{subfigure} 
\usepackage{enumerate}
\usepackage{amssymb}

\begin{document}

\title{Distributed Power Control Network and Green Building Test-bed for Demand Response in Smart Grid}

\author{\IEEEauthorblockN{Kei Sakaguchi \IEEEauthorrefmark{1}  \IEEEauthorrefmark{2},
Van Ky Nguyen \IEEEauthorrefmark{2},
Yu Tao \IEEEauthorrefmark{2}\\,
Gia Khanh Tran \IEEEauthorrefmark{2},
Kiyomichi Araki \IEEEauthorrefmark{2}} \\
\IEEEauthorblockA{\IEEEauthorrefmark{1}Osaka University, Osaka, Japan\\
Email: \{sakaguchi\}@comm.eng.osaka-u.ac.jp}\\
\IEEEauthorblockA{\IEEEauthorrefmark{2}Tokyo Institute of Technology, Tokyo, Japan\\
Email: \{sakaguchi, ky, yutao, khanhtg, araki\}@mobile.ee.titech.ac.jp}}

\maketitle

\begin{abstract}
It is known that demand and supply power balancing is an essential method to operate power delivery system and prevent blackouts caused by power shortage. In this paper, we focus on the implementation of demand response strategy to save power during peak hours by using Smart Grid. It is obviously impractical with centralized power control network to realize the real-time control performance, where a single central controller measures the huge metering data and sends control command back to all customers. For that purpose, we propose a new architecture of hierarchical distributed power control network which is scalable regardless of the network size. The sub-controllers are introduced to partition the large system into smaller distributed clusters where low-latency local feedback power control loops are conducted to guarantee control stability. Furthermore, sub-controllers are stacked up in an hierarchical manner such that data are fed back layer-by-layer in the inbound while in the outbound control responses are decentralized in each local sub-controller for realizing the global objectives. Numerical simulations in a realistic scenario of up to 5000 consumers show the effectiveness of the proposed scheme to achieve a desired $10\%$ peak power saving by using off-the-shelf wireless devices with IEEE802.15.4g standard. In addition, a small scale power control system for green building test-bed is implemented to demonstrate the potential use of the proposed scheme for power saving in real life.

\end{abstract}

\section{Introduction}\label{intro}

After the nuclear accidents at Fukushima Daiichi nuclear power plant due to the terrible earthquake and tsunami which hit the Tohoku areas in Japan on 11th of March 2011, Japan has faced a serious shortage of power supply especially in Tokyo area. Following efforts to prevent the same accidents, Japan has stopped almost all of its nuclear power plants so far. This leads to a serious shortage of power supply in all land of Japan now and in the near future. As a result, the load power consumed during peak hours should be reduced by 10$\% $ to 15$\%$ as stated by the government. To avoid tremendous power blackout over the whole area due to power overload, area-by-area planned blackouts were conducted in Tokyo in 2011 and are planned to be conducted again in Osaka and some areas of Japan this year. However, the cost of such planned blackouts on our daily life and business is prohibitive, and in a long-term more radical solutions on this power saving issue should be more considered. In this context, as the vision of future electric power system integrated with Information Communication Technology (ICT) to solve the energy problems, new power system called Smart Grid\cite{intro1} has attracted much attentions recently.

\begin{figure} [h]
\centering
\includegraphics[width=8.5cm]{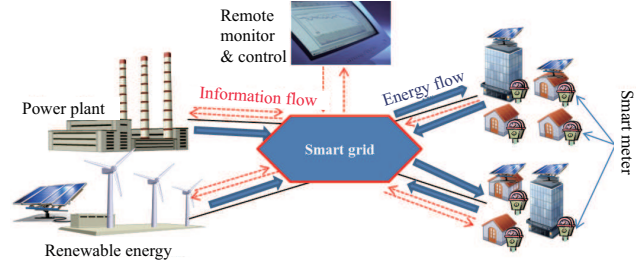} 
\caption{Concept of smart grid.}
\label{sg} 
\end{figure}

Figure\ \ref{sg} shows the concept of Smart Grid. In Smart Grid, there are many entities such as conventional centralized big-size power generator including nuclear power plant, geographically distributed small-size power generators such as solar and wind energy, distributed energy storage devices including heat transformer, end-user devices like industrial plant and home appliance, and in-the-middle distribution power line network including power transfers. Futhermore, beside the physical connection of power line, a communication network connecting all these entities is also a part of Smart Grid which can help to improve operability of power systems. Especially, knowledge about power consumption at the consumer side can be collected by using such sensor like devices called smart meter. Under the described framework of Smart Grid, Demand Response (DR) which can efficiently control the power consumption of consumers regarding the available power supply is the key technology in the implementation of future Smart Grid. 

\begin{figure} [h]
\centering
\includegraphics[width=8.5cm]{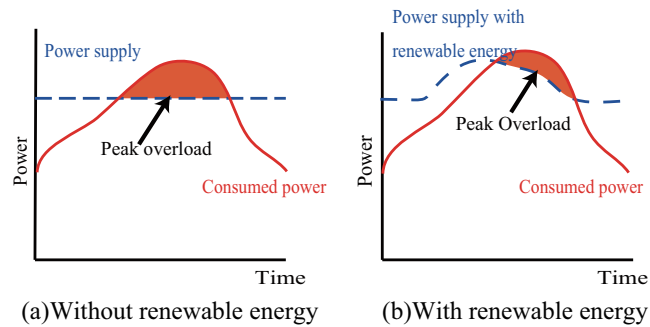} 
\caption{Demand and supply power balance.}
\label{DR} 
\end{figure}

In general, DR service is defined as a dynamic mechanism to manage the electricity power consumption of consumers in response to the supply condition in the whole system \cite{Intro2b},\cite{Intro3} as illustrated in Fig.\ \ref{DR}. As seen in Fig.\ \ref{DR}(a), if the potential peak power demand is larger than the maximal power supply, the DR service should be applied to customers to decrease the power overload to be lower than the supply power. In addition, some kinds of time-varying renewable energy sources such as solar or wind can be utilized to increase the power supply capacity as in Fig.\ \ref{DR}(b), however the gap between demand and supply power still exists, which degrades the quality of the power system. Hence, the DR needs to be applied to both the customers and the utilities such as energy storage devices for demand-and-supply power balancing. For both cases, the incentive metering information of customers are needed frequently and the control response should be sent back to the customers if there is a demand-supply gap, which can be seen as a  feedback loop of power control network. However, the communication architecture of such kind of control network is still an open problem. The design of the network architecture obviously depends on the required response speed of the control. In the DR case, the response speed should be very high due to the rapidly changing power demand during peak hours, especially in the large power system such as metropolitan electricity system. Hence, the design of network architecture for this  control network is quite important to realize DR for peak load power shaping \cite{Intro4}\cite{Intro5}.

As reported in \cite{Intro6} ,\cite{Intro7}, a theoretical bound of maximal allowable delay or minimum rate of communication was derived to guarantee stable control via a point to point communication network, but there was no discussion about designing of point to multi-point network  which can be applied for large-scale system such as Smart Grid. Authors in \cite{Intro2a} proposed how to design a smart message authentication system to reduce control delay time during peak hours, however the DR was not included. On the other hand,  \cite{Intro2b} evaluated the performance of DR restricted in a smale-scale Home Energy Manegement System, or commercially called home-gateway, with negligible delay. However, delay becomes the most critical parameter in realizing stable control in large-scale feedback systems like Smart Grid. Implementing a simple star-typed centralized architecture consisting of a central controller and multiple sensors and actuators, since feedback delay increases with respect to the size of the network, leads to an unstable control as reported in \cite{Intro8},\cite{Intro9}. In addition, coverage of the network becomes limited in the case of wireless access in the centralized architecture. Although multi-hop architecture \cite{KyLeng} can solve the problem of restricted coverage, it worsens feedback performance due to packet relay. Therefore, instead of the centralized control architecture, a distributed architecture with multiple controllers is a reasonable solution for large-scale networks. However, to the best of our knowledge its design criteria are not yet established.

This paper establishes an hierarchical distributed power control network achieving control stability even in large-scale networks. In the proposed architecture, sub-controllers are introduced to partition the network into smaller clusters, in which corresponding local feedback loops are generated with shorter delays to achieve stable control. At the same time, the global objective of demand-and-supply power balancing can still be realized through vertically exchanging feedback information and control responses between sub-controllers in different layers. If the size of each local feedback loop is kept to satisfy stable control condition, the control network becomes scalable regardless of the number of houses or actuators by appropriately partitioning the network into stable control clusters and stacking them into layers in an hierarchical manner. We set up numerical simulation in a scenario of 5000 houses to show the validity of the proposed hierarchical architecture in realizing peak load power reduction, while the performance of the centralized architecture becomes unstable. Finally, a green building test-bed \cite{Ky1},\cite{Ky2} employing the proposed control scheme was developed and its power saving performance with respect to varying demand power is validated. Although, the test-bed is a small-scale model\cite{Ky1},\cite{Ky2}, the results are still valuable for large-scale networks owing to the scalability of the proposed algorithm.

The rest of this paper is organized as follows. Section 2 describes the architecture and algorithm of the proposed distributed power control network. Section 3 presents numerical simulation results to validate the proposed scheme. Furthermore, a green building test-bed employing the proposed distributed power control network is introduced in Sec. 4. Finally, the conclusion is given in Sec. 5.

\section{Distributed Power Control Network and Algorithm}
\label{sec2}

\subsection{Hierarchical distributed power control network}

\begin{figure}[h] 
\centering
\includegraphics[width=8.5cm]{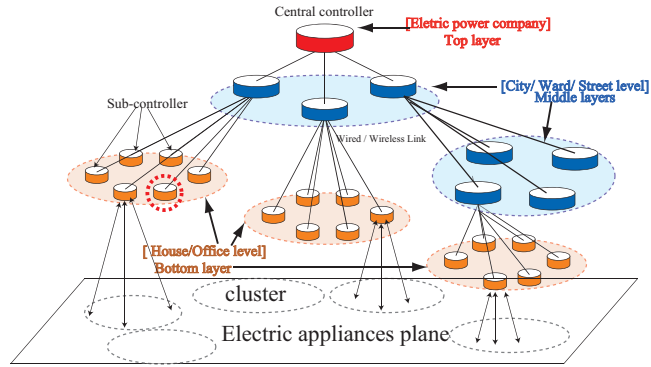} 
\caption{Hierarchical distributed control network.}
\label{Hierarchical} 
\end{figure}

In DR service, the knowledge of power consumption of the whole system is required in making control decision. Generally, there are two measures of calculating the consumed power in the centralized control system. The conventional measure is to monitor the output power from electric power generators or distribution lines, however, this scheme cannot identify the electric appliances being used by customers. As the result, the DR cannot apply the control to the customer side. The other measure is to calculate the summation of all consumed power by monitoring all the consumed power of electric appliances. By this measure, the knowlegde of electric appliances being used can be known, however, the time spent for gathering data is prohibitive since it is almost proportional to the number of appliances. This means that the control performance based on this measure becomes unstable especially in the power systems with a large number of houses and appliances. To solve the problem of unstable control in such large systems, we propose an hierarchical distributed power control network shown in Fig.\ \ref{Hierarchical}. The control network consists of bottom layer sub-controllers, middle layer sub-controllers and top layer central controller corresponding to specific control functions. The role of the sub-controller in each layer is explained as follows.

\begin{itemize}
\item Sub-controllers at bottom layer: partition the large system into many small-size clusters with local feedback loops. Here, each sub-controller can directly change the power consumed by electric appliances adapting to its assigned power consumption limit. The control in the cluster becomes stable if the cluster size is small enough to keep the control latency below a maximal allowable duration. 
\item Sub-controllers at middle layers: feedback information of power consumption from lower layer to upper layer and determine the power consumption limits of the sub-controllers at their subsequent lower layer, by which the power consumption of the clusters at the bottom layer can be adjusted indirectly. 
\item Central controller at top layer: keeps demand-and-supply power balancing by comparing the overall power consumption and the maximal supply power from energy generators to assign appropriate power consumption limits to its sub-controllers at lower layer.
\end{itemize}

It is obvious that the global objective of DR control is realized through assigning power consumption limits of sub-controllers turn by turn. The power consumption limits are periodically sent to the sub-controllers at bottom layer. During this interval, the local sub-controllers at bottom layer can simultaneously adjust the power consumption of their cluster to fulfill the assigned limit without wasting the time in waiting for responses from other clusters. In the next sub-section, the control algorithm for sub-controllers at each layer will be explained.

\subsection{Control Algorithm}
First of all, we explain the notations used in this section of the paper. For a specific sub-controller $i$ of layer $l$ ($l = 1$ for the top layer and $l=L$ for the bottom layer), $P^{\rm c}_{l,i}$ denotes the total consumption power of all electric appliances under the control of this sub-controller. ${P}^{\rm s}_{l,i}$ denotes the power consumption limit assigned to this sub-controller. Since there is only one sub-controller at the top layer, $P^{\rm c}_{1,1} = P^{\rm c}$ represents the total consumption power of the system and ${P}^{\rm s}_{1,1} = {P}^{\rm s}$ is the power consumption limit of the whole system, which equals the total power supply in this paper. On the other hand, the power consumption of the electric appliance $k$ is denoted as $P^{\rm c}_{\textrm{e},k}$.

Since this paper focuses on power saving during peak hours and discrete time control is assumed, the global objective can be described as a function of discrete time  inequality below, 

\begin{eqnarray}
 P^\mathrm{c}(kt_0) \lesssim P^\mathrm{s}(kt_0), 
\label{eq1}                           
\end{eqnarray}
where $\lesssim$ denotes the control operator of suppressing power below an assigned value and $t_0$ is the time interval of making decision on power control.

\begin{figure*} [t]
\centering
\includegraphics[width=16
cm]{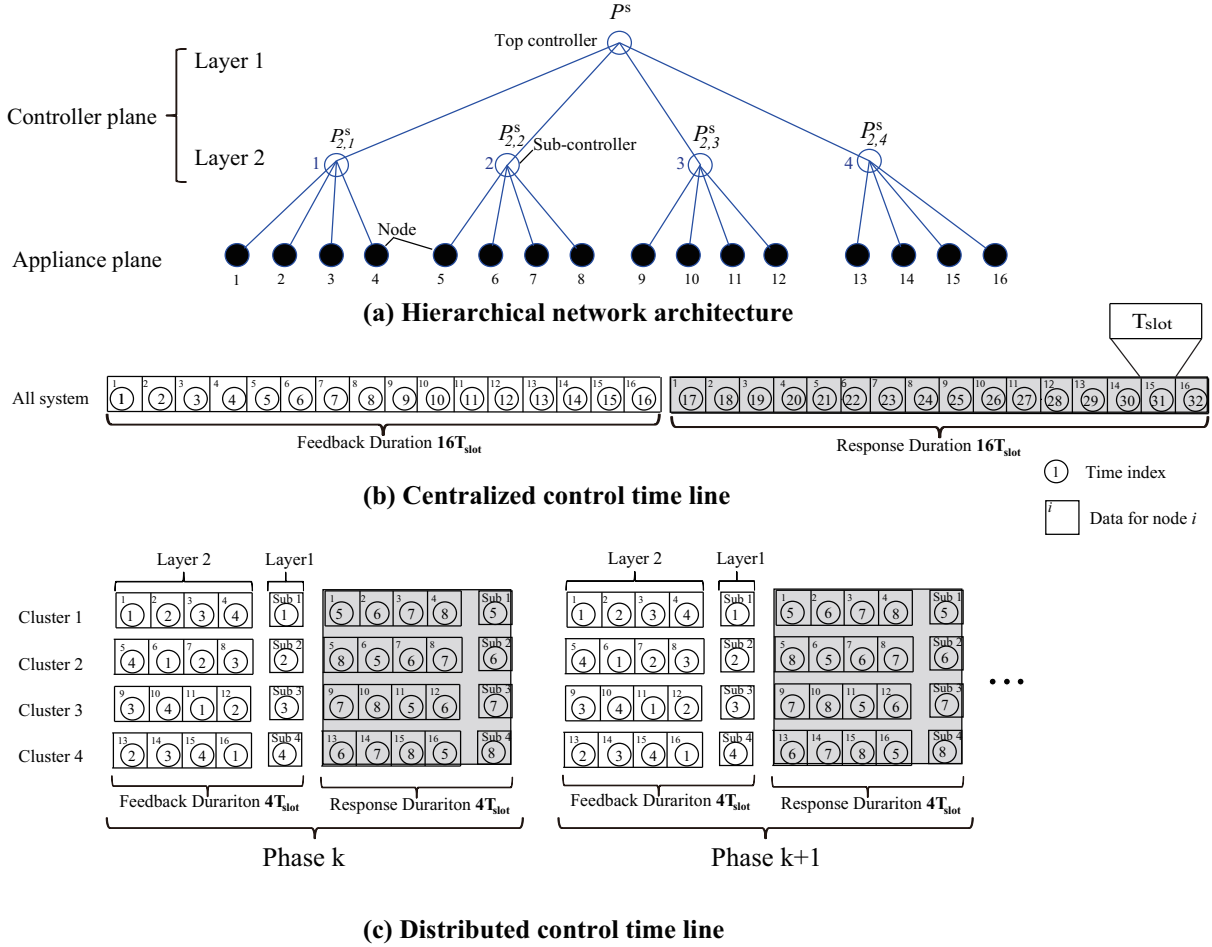} 
\caption{An example proposed network and communication framework.}
\label{Algo1} 
\end{figure*}

Let us assume a power control system consisting of a total of $N$ electric appliances. In the case of centralized control, all the power consumption data are required to be collected prior to decision making of control response. Therefore, $t_o$ turns into $2\times N \times T_\mathrm{slot}$  assuming a simple Round Robin fashion as described in Fig. \ref{Algo1}(b), where $T_\mathrm{slot}$ is the length of time slot assigned for each uplink (feedback) and downlink (response) transmission. For such a long interval, the consumed power might be increased above the limit with high probability, which leads to unstable DR performance. 

In contrast, the distributed control with multiple sub-controllers can keep the consumption power under the limit by partitioning the system into small clusters to reduce the control time. 

\subsubsection{Simple scenario of distributed control algorithm}
In order to explain the distributed control algorithm, a simple example with $N = 16$ electric appliances is illustrated in details before describing the general scheme. Figure \ref{Algo1} shows the structure of the control network in this simple scenario.

In Fig. \ref{Algo1}(a), the control network forms a 4-ary tree topology where each parent node has four children nodes. The terminal vertices (black circles) of the tree correspond to electric appliances, which form a so-called electric appliance plane in this paper. All the other vertices denote sub-controllers and they form a controller plane. The degree of each sub-controller is defined by the number of children nodes connected to the sub-controller. Especially, the degree of a sub-controller at the bottom layer of the controller plane or the number of electrical appliances grouped by that sub-controller is called as cluster size $N_{\rm cs}$ and is the most important parameter in this paper. For simple illustration, the degrees of all sub-controllers are the same as $N_{\rm cs} = 4$. Within the controller plane, the tree is separated into $L$ layers where the root is called the top layer or layer 1 and the layer $L$ is the bottom layer. The number of layers in this case are determined by $L = \log_{N_{\rm cs}}N = \log_{4}16 = 2$. On the other hand, the electric appliances are grouped into four clusters, each of which is controlled by a sub-controller at layer 2 (or called home-gateways in commercial). These four sub-controllers at layer 2 are then again controlled by a central controller at layer 1 (top layer). In this paper, a sub-controller which is controlled by another sub-controller at its upper layer is called the children sub-controller of the upper layered one. Inversely, a sub-controller which supervises another sub-controller at its lower layer is called the parent sub-controller of the lower layered one.

As electric appliances are partitioned into four different clusters, the time period of the control loop can be reduced into $t_0 = 2 \times N_{\rm cs} \times  T_{\rm slot}$ as shown in Fig.~\ref{Algo1}(c) owing to parallel processing of each cluster and layer. For example, at the time index of 1, four appliances numbered \{1, 6, 11, 16\} and the sub-controller 1 at layer 2 simultaneously communicate with their corresponding parent sub-controllers. It is noted that there is only one appliance can communicate every time slot in the case of centralized control.

In the initial phase, the power consumption limit is set equally as ${P}^{\rm s}_{2,i} = P^{s}/4$. After that, the control algorithm is repeated every interval of $t_0$ at each layer of the control system as follows. 

\paragraph{At the bottom layer (layer 2):}
At the bottom layer, each sub-controller has a list of electric appliances and their corresponding power usage priority. Without loss of generality, we assume electric appliance with lower priority has lower device index. The function of these sub-controllers at bottom layer is to appropriately turn off low priority electric appliances such that the power consumption of the cluster after control satisfies the power consumption limit assigned by the parent sub-controller at layer 1. The algorithm is summarized using mathematical formulas for a specific bottom layer sub-controller with index $i$ as follows. 
\begin{eqnarray}
& &  N_{\rm off} =  \arg\min_{N_{\rm off} \geq 0} \left( {P}^{\rm s}_{2,i}  -  \sum_{ch(i)=N_{\rm off}+1}^{N_{\rm cs}} P^{\rm c}_{\textrm{e},ch(i)}\right)^2 \nonumber \\
& & \textrm{subject \ to}   \nonumber\\
& & \sum_{ch(i)=N_{\rm off}+1}^{N_{\rm cs}} P^{\rm c}_{\textrm{e},ch(i)}  \leq {P}^{\rm s}_{2,i}  \nonumber\\
& & \ P^{\rm s}_{\textrm{e},ch(i)} := 0 \quad  \forall ch(i)  \leq N_{\rm off} \quad \textrm{(power turn off)}\nonumber\\
& & \ P^{\rm s}_{\textrm{e},ch(i)} := P^{\rm c}_{\textrm{e},ch(i)} \quad  \forall ch(i)  > N_{\rm off} \quad \textrm{(power maintain)}\nonumber
\end{eqnarray}
where $ch(i)$ denotes the children of node $i$ and $P_{{\rm e},ch(i)}^{\rm c}$ denotes the consumed power of a child electric appliance controlled by sub-controller $i$. Here, $P_{{\rm e},ch(i)}^{\rm c}$ is measured at the electric appliance $ch(i)$ and fed back to its parent sub-controller $i$ using the assigned feedback time slot as depicted in Fig.~\ref{Algo1}(c). After achieving feedback information about the power consumption of its children electric appliances, the sub-controller calculates the total power consumption of the cluster $P^{\rm c}_{2,i} = \sum_{ch(i)}P^{\rm c}_{\textrm{e},ch(i)}$ and confirms if the value does not exceed its pre-assigned power consumption limit ${P}^{\rm s}_{2,i}$. Otherwise, the sub-controller finds the smallest number of low priority electric appliances to turn off such that the power consumption condition holds. Then, it responses a turn-off command \{$P^{\rm s}_{\textrm{e},ch(i)}$\} to the corresponding selected low priority electric appliances. During normal hours of a day when there is no power overload, the downlink traffic from sub-controller to its electric appliances is rather light. Thus these communication resources can be saved for other purposes. During peak hours, the sub-controller needs to select a positive number $0 < N_{\rm off} < N_{\rm cs}$ of devices to turn off. 

\paragraph{At the top layer (layer 1):}
The sub-controller at top layer (central controller) determines the power consumption limits ${P}^{\rm s}_{2,i}$ of the sub-controllers at bottom layer using the following algorithm,
\begin{eqnarray}
& & \{{P}^{\rm s}_{2,1}, \dots,{P}^{\rm s}_{2,N_{\rm cs}}\} = \arg\min_{\{{P}^{\mathrm{s}}_{2,i}\}} \sum_{i=1}^{N_{\rm cs}} \Bigl( {P}_{2,i}^{\mathrm{s}}-P_{2,i}^{\mathrm{c}}\Bigr)^2  \nonumber \\
 & & \rm{subject \ to} \nonumber\\
 & & \sum_{i=1}^{N_{\rm cs}}{{P}_{2,i}^{\mathrm{s}}}={P}^\mathrm{s}. \nonumber
\end{eqnarray}
In this algorithm, based on the feedback information about power consumption  $P^{\rm c}_{2,i}$ from its children sub-controller at layer 2, the central controller calculates the corresponding power consumption limits $P^{\rm s}_{2,i}$ to achieve demand-and-supply power balancing by finding their minimum mean square distance under the constraint of the available power supply $P^{\rm s}$. After the optimized power consumption limits  $P^{\rm s}_{2,i}$ are determined, these values are sent back to the sub-controllers at layer 2. The sub-controllers then use these power consumption limits for decision making in the next iteration of the control process. During peak hours with power excess, if the electric power consumption of different clusters is uniformly distributed, i.e. $P^{\rm c}_{2,i} = P^{\rm c}_{2,i^\prime} \ \forall i \neq i^\prime$, the optimized power consumption limits are then $P^{\rm s}_{2,i} = P^{\rm s}/N_{\rm cs}$. In other words, the central controller attempts to distribute its available power to its uniform clusters equally. On the contrary, when the power distribution between different clusters is non-uniform, this algorithm still guarantees that the cluster consuming more power is assigned with a higher value of power consumption limit under the sum power consumption constraint. Therefore, the power allocation introduced in this paper is spontaneously strict in terms of power suppression and fairness in terms of power allocation toward clusters.   


Owing to the above algorithm, the power system can be controlled with a shorter control period, which leads to better performance. There might be different methods to determine such power consumption limit values, however the proposed algorithm is selected since it gives unique solution of limit values by finding the minimum mean square distance between demand and supply power. 

\subsubsection{General algorithm}
Following the explanation for the simple example with 16 electric appliances, we generalize the algorithm for an arbitrary number of $N$. For general purpose, we consider the scenario that $N$ electric appliances are grouped into $N_{\rm g}$ clusters with size $N_{\rm cs} = \lceil N/N_{\rm g}\rceil$.  As explained previously, the controller plane of the distributed control network is then formed as a tree topology with $L = \lceil \log_{D}N_{\rm g}\rceil+1$ layers of sub-controllers where $D$ denotes the degree of each sub-controllers at middle and top layers. The algorithms explained in previous subsection can be extended for arbitrary number of layers as follows. 

\paragraph{At the bottom layer (layer $L$):}
At the bottom layer, each sub-controller has a list of electric appliances and their corresponding power usage priority. The function of these sub-controllers at bottom layer is to appropriately turn off low priority electric appliances such that the power consumption of the cluster after control satisfies the power consumption limits assigned by the parent sub-controller at layer $L-1$. The control algorithm of a sub-controller at bottom layer is similar as that in the simple example thus detailed explanation is omitted for brevity. However, different from the simple example with the cluster size of four, it should be noted that $N_{\rm cs}$ in real scenario has the order of hundreds or thousands of electric appliances.

\begin{figure} [h]
\centering
\includegraphics[width=8.5cm]{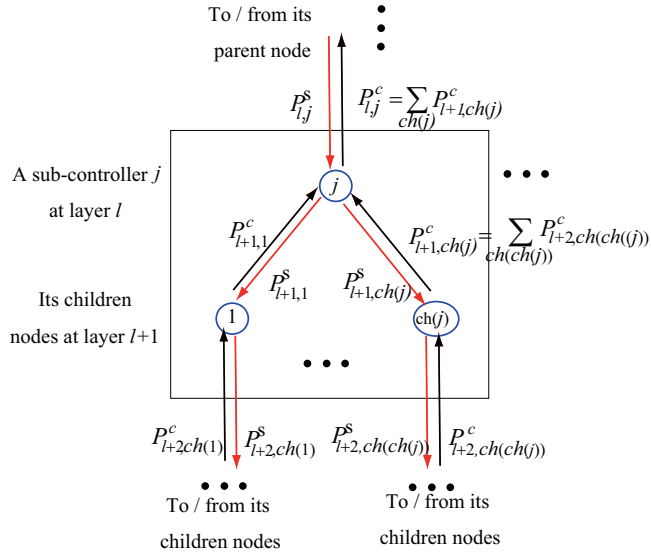} 
\caption{Distributed feedback and response mechanism.}
\label{Algo2} 
\end{figure}

\paragraph{At the middle layer and the top layer:}
As depicted in Fig.~\ref{Algo2}, a sub-controller $j$ at a layer $l \neq L$ determines the power consumption limits ${P}^{\rm s}_{l+1,ch(j)}$ of its children sub-controller at the lower layer $l+1$ using the following algorithm.
\begin{eqnarray}
& & \{{P}^{\rm s}_{l+1,ch(j)}\} = \arg\min_{{P}^{\mathrm{s}}_{l+1,ch(j)}} \sum_{ch(j)=1}^{D} \Bigl( {P}_{l+1,ch(j)}^{\mathrm{s}}-P_{l+1,ch(j)}^{\mathrm{c}}\Bigr)^2 \nonumber \\
 & & \rm{subject \ to}  \nonumber\\
 & & \sum_{ch(j)=1}^{D}{{P}_{l+1,ch(j)}^{\mathrm{s}}}={P}^\mathrm{s}_{l,j} \nonumber
\end{eqnarray}
Here, similarly as the simple example, based on the feedback information about power consumption  $P^{\rm c}_{l+1,ch(j)}$ from its children sub-controllers $ch(j)$ at layer $l+1$, the sub-controller $j$ calculates the corresponding power consumption limits $P^{\rm s}_{l+1,ch(j)}$ to balance the demand and supply of power under its assigned constraint of consumption power $P^{\rm s}_{l,j}$. It is noted that $P_{l+1,ch(j)}^{\mathrm{c}} = \sum_{ch(ch(j))} P_{l+2,ch(ch(j))}^{\rm c}$ is calculated at the sub-controller $ch(j)$ at layer $l+1$ by summing the power consumption information sent from its children sub-controllers $ch(ch(j))$ at layer $l+2$.  After the optimized power consumption limits  $P^{\rm s}_{l+1,ch(j)}$ are determined, these values are sent back to the children sub-controllers of $j$ at layer $l+1$. The children sub-controllers then use these power consumption limits for decision making in the next iteration of the control process.  

It is remarkable that the above algorithm is general at any middle and top layers of the distributed control network. At each corresponding sub-controller, as mentioned in the simple case, the algorithm guarantees that the power constraint is satisfied while available power is distributed proportionally to the demands of the children sub-groups. Therefore, the algorithm is scalable regardless of the number of electric appliances in terms of guaranteeing peak power suppression, even for the real scenario with the number of electric appliances up to the order of million nodes. However, to realize such large-scale distributed control networks, it is important to design the cluster size $N_{\rm cs}$ and the sub-controller degree $D$ in accordance to the communication link capacity. As this paper assumes a perfect backbone network for communication between sub-controllers, it only focuses on the design of $N_{\rm cs}$.

\subsection{Iterative control algorithm and home-gateway}

The distributed algorithm explained in Sec.2.2 is baseline, and there are many ways to improve control performance by introducing some approximation and a priori knowledge of electric appliances. As such schemes, we introduce an iterative control algorithm and the concept of home-gateway in this paper.

\subsubsection{Iterative control algorithm} 
	In the baseline algorithm, a sub-controller receives all feedback and sends back response afterward to all its children sub-controllers or appliances. In the iterative control algorithm, this order of feedback and response is changed as child sub-controller by child sub-controller or appliance by appliance. As such the waiting time per child node can be reduced to $t_o$ while the control period per cluster remains the same. It is noted that the consumed power at the mother sub-controller should be calculated iteratively in this algorithm. Since the response is sent before waiting all the feedback, this algorithm is considered as an approximation of the baseline algorithm.
\subsubsection{Home-gateway}
  
  As mentioned in Sec.1, there are varieties of entities and electric appliances in Smart Grid, thus it is quite complex to monitor and control the whole large-scale power systems. However, it is neccesary to monitor all the electric appliances to identify which appliances being used and calculate the total consumption power. It is clearly that monitoring all the appliances one by one makes the control time become long. To reduce the time spent for monitoring the appliances, smart metering devices called home-gateway can be introduced. Home-gateway can calculate the total power consumption in the house by directly measuring the output power from power line, and it can further analyze the shapes of electric current and voltage to identify which appliances are being used as reported in \cite{kato}. By using the home-gateway, the consumed power at several appliances can be calculated with quite small latency.

\begin{figure*} [t]
\centering
\includegraphics[width=15
cm]{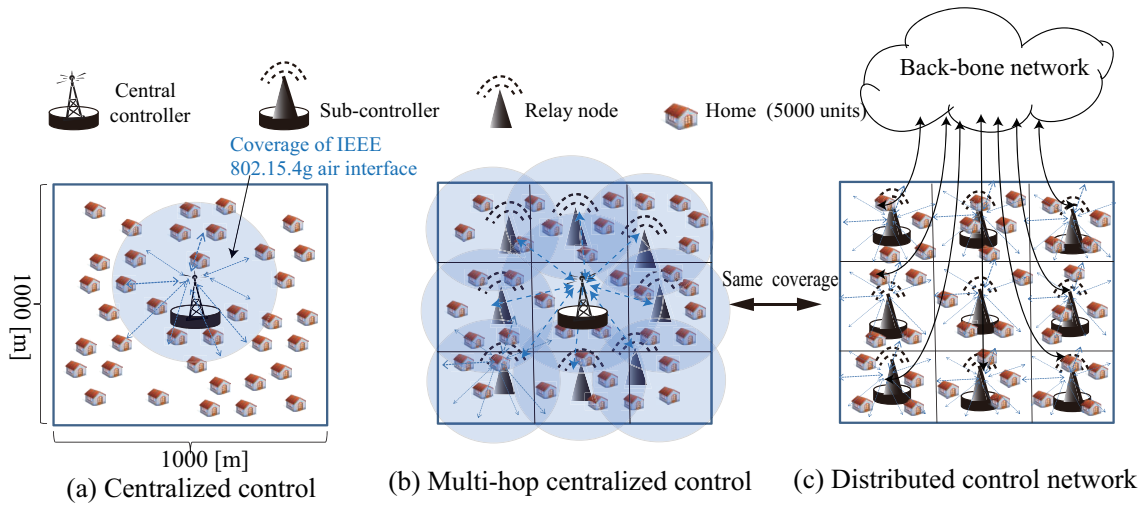} 
\caption{Target scenario for simulation.}
\label{TS} 
\end{figure*}

\section{Numerical Analysis}

In this section, in order to both evaluate the performance of the proposed control network and explain how to specify the parameters to design the control network,  numerical simulations in a realistic scenario are conducted and taken into discussion.

\subsection{Target Scenario}
The target scenario is illustrated in Fig.\ \ref{TS}. There is an area of 1 $\mbox{km}^2$ with up to 5000 houses sharing the same electric power line. As reported in \cite{Intro10}, an average peak load power per house of 1200W is assumed in this scenario as the representative value in the case of Japan, so that in total 6MW is usually consumed  during peak hours. If the local power generator only supplies the maximum power lower than 6MW, blackout will occur in this area during peak hours. To prevent the blackout, a smart grid with centralized cluster controller and home-gateways at every house are introduced to reduce the peak power consumption $10\%$ lower than the maximum supply power. At each house, the home-gateway classifies the home appliances into two groups of high and low priorities. High priority appliances  are prerequisite for normal life and cannot be turned off, while low priority appliances can be turned off in a predefined order to save power during peak hours. Several typical electric appliances including air-conditioner (831W/unit) and refrigerator (268W/unit) with high prioprity, lamp (60W/unit) and television (141W/unit) with low prioprity, in each house are emulated to produce an average peak load power of 1200W in a duration of one hour. Besides, the On/Off switches of all appliances are generated based on a random Poisson process to imitate variable load power.  The details of simulation parameters are shown in Table I. 

As illustrated in Fig.\ \ref{TS}, there are three control network topologies used in this scenario: a star-typed centralized control topology (Fig.\ \ref{TS}a), a star-typed centralized control  topology with multi-hop (Fig.\ \ref{TS}b) and a proposed distributed control topology (Fig.\ \ref{TS}c). In all cases, the standard wireless protocol of IEEE802.15.4g for smart meter is used in wireless link between home-gateway and the cluster controller, where the time slots are allocated equally for the downlink and uplink. Moreover, to ensure fairness of control among different houses in the clusters, guaranteed time slot of IEEE802.15.4g is assigned for each home-gateway using simple Round Robin scheduling. In the star-typed centralized control case, it is obvious that the coverage of wireless link is limited to the houses near the central controller. By using multi-hop transmission with relay devices for smart meter, the wireless coverage can be extended but transmission time also becomes larger, which leads to longer control delay. In the distributed control network, it is assumed that sub-controllers are connected to the upper layer controller via a backbone network such as Power Line Communication (PLC) with a moderate rate up to 1Mbps. Each cluster is assumed to be assigned with an orthogonal channel not interfered by other clusters.

\begin{table} [h]
\renewcommand{\arraystretch}{1.3}
\caption{Simulation parameters.}
\label{SimPar} 
\centering
\begin{tabular}{l || c}
\hline			
  Parameter & Value \\
\hline		
Target area & $1000\mbox{m}^2$ \\	
Number of houses/consumers & $500\sim 5000$  \\ 
Average peak power per house & 1200W \\
Average limit power per house & 1080W \\
Air interface & IEEE802.15.4g 50kbps\\
Transmit power & 250mW\\
Channe model & Rayleigh fading \\
Path-loss exponential & 3.2\\
Shadowing variance & 8dB\\
Modulation scheme & Binary FSK\\
Coding rate & 1/2\\
Packet length & 32bytes\\
Time slot length & 960 symbols \\
 \hline  
\end{tabular}
\end{table}

\subsection{Simulation Results}

\begin{figure} [h]
\centering
\includegraphics[width=8.5cm]{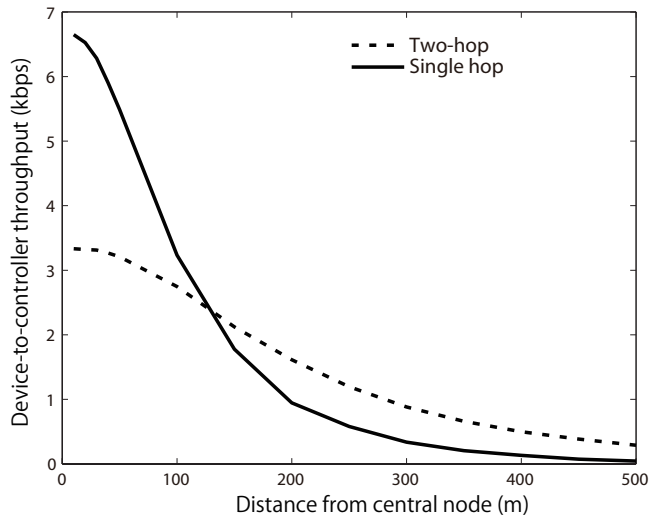} 
\caption{Average throughput vs. distance.}
\label{Thr} 
\end{figure}

\begin{figure} [h]
\centering
\includegraphics[width=8.5cm]{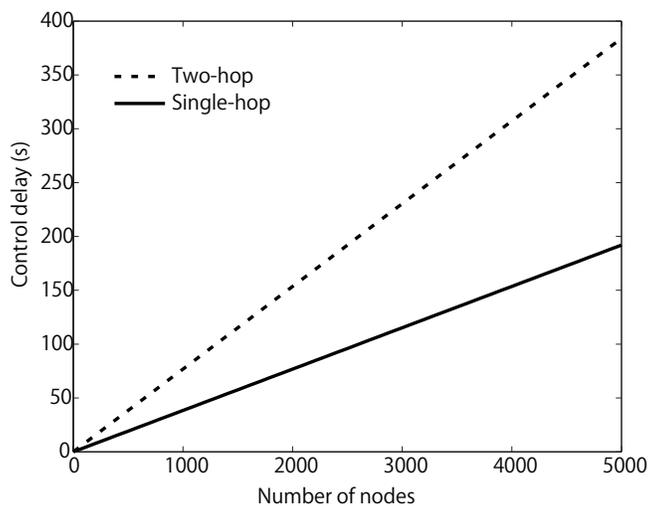} 
\caption{Average control delay vs. number of nodes.}
\label{Conper} 
\end{figure}

 Figure\ \ref{Thr} shows the average throuhput at controller versus the distance between a house and the controller in the case of star-typed network with and without multi-hop transmission, which uses one relay device ideally placed in the middle between the house and the controller. The throughput is calculated by counting the number of correctly received packets at controller when the houses transmit packets in the TDD fashion with IEEE802.15.4g air interface. The maximum throughput is approximately $50\mbox{kbps}\times \frac{32\times8}{960\times2}\approx 7\mbox{kbps}$. Figure\ \ref{Thr} shows that the throughput decreases with the distance in both cases, however using multi-hop transmission the coverage can be extended to the farthest houses at 500m distance. On the other hand, Fig.\ \ref{Conper} shows the period required for control of each home-gateway with respect to the number of houses in the same network topologies. It is obvious that the time spent for multi-hop transmission is much longer than that of the direct transmission, which is not preferable to control system.

\begin{figure} [h]
\centering
\includegraphics[width=8.5cm]{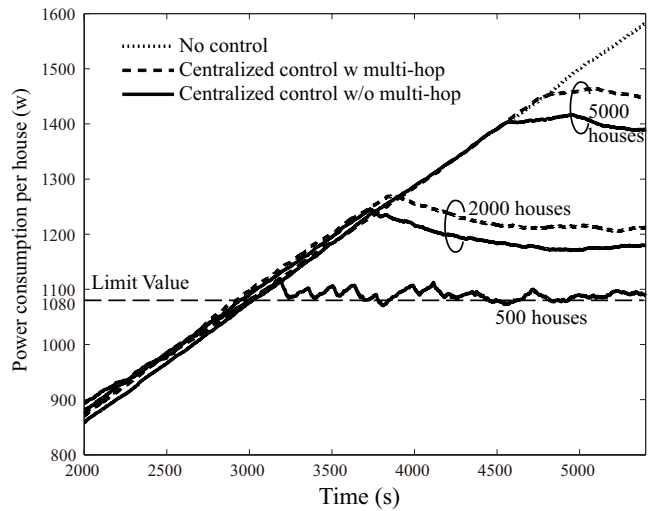} 
\caption{Power control performance of centralized scheme.}
\label{Cen} 
\end{figure}

   Figure \ref{Cen}  shows the simulation results on the control performance in the case of using star-typed centralized control network with and without multi-hop transmission. It reveals that the centralized control scheme can only fulfill the objective limit of power consumption when the number of houses is less than 500 and the control performance degrades dramatically as the number of houses increases. Because the time required for gathering feedback data from all the home-gateways in the area increases with the number of houses, the larger number of houses results in the larger control latency, that yields a larger amount of power overloaded. One more reason is that the number of home-gateways at the edge of transmission coverage also increases. By using 2-hop relay transmission, the number of uncontrollable home-gateways decreases owing to a larger area of coverage. However, the longer control period spent for multi-hop transmission leads to an even degraded control performance in terms of suppressing the power consumption.
   
\begin{figure} [h]
\centering
\includegraphics[width=8.5cm]{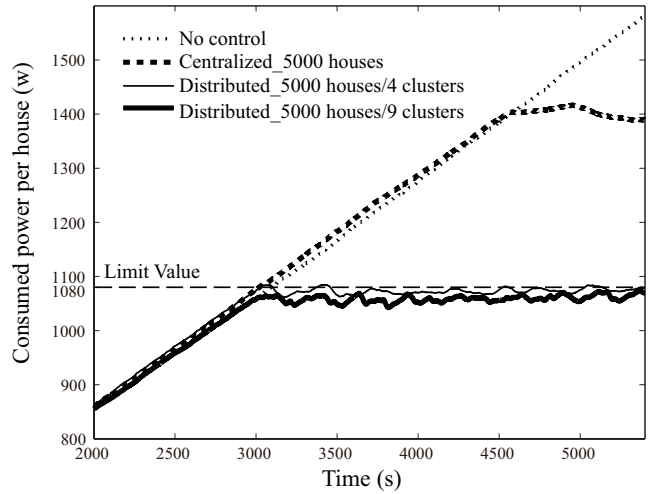} 
\caption{Power control performance of distributed scheme.}
\label{Dis} 
\end{figure}

\begin{figure} [h]
\centering
\includegraphics[width=8.5cm]{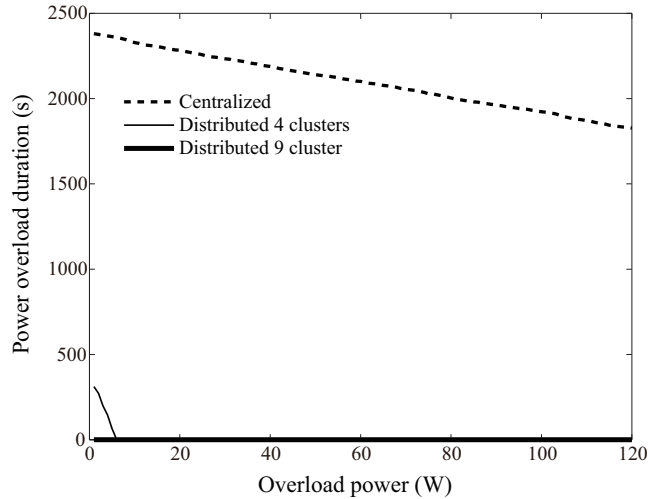} 
\caption{Overload power performance.}
\label{Ove} 
\end{figure}

\begin{figure} [h]
\centering
\includegraphics[width=8.5cm]{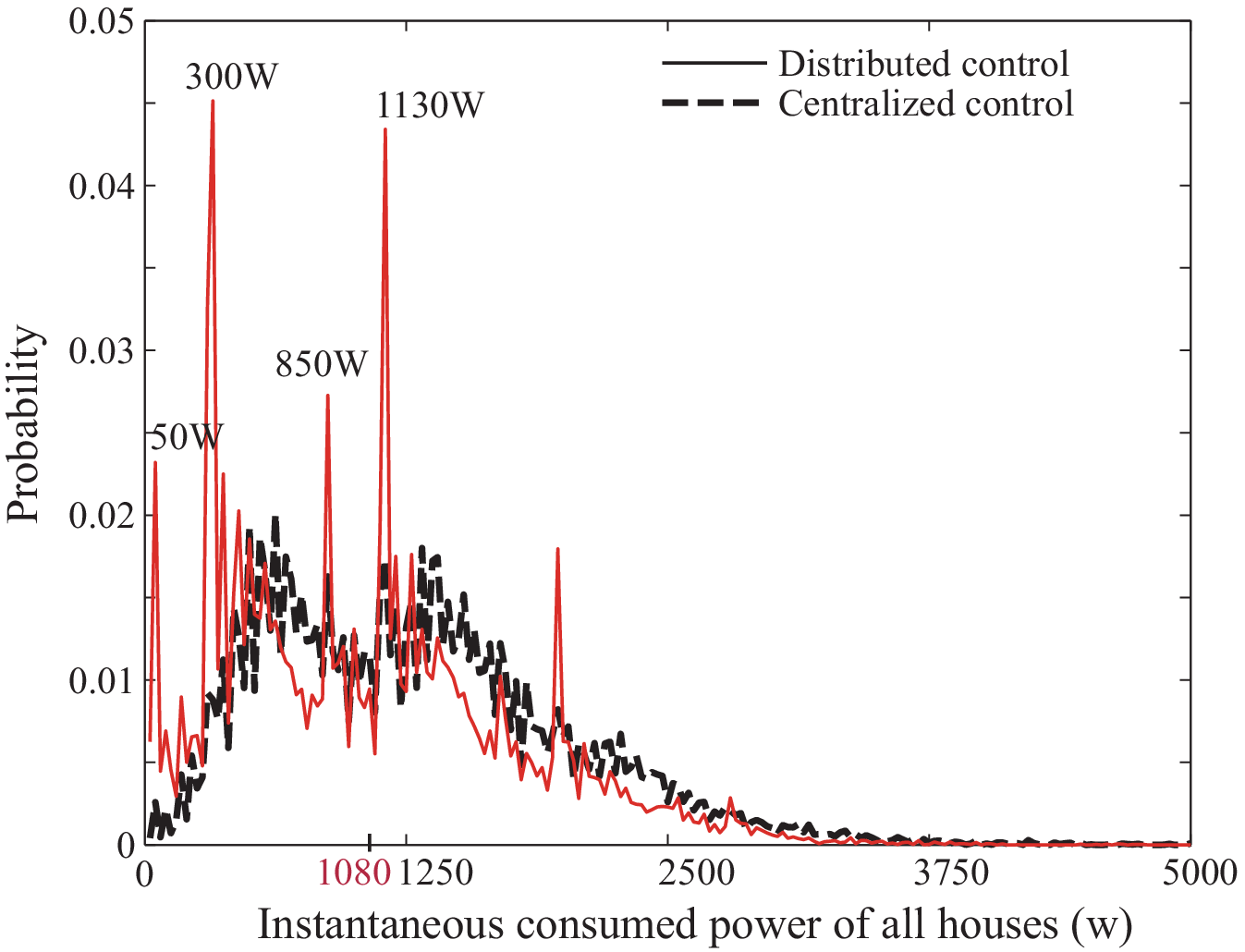} 
\caption{Probability distribution of instantaneous consumed power.}
\label{powerdistribution} 
\end{figure}

The results on control performance of the distributed power control network with two layers are shown in Figs.\ \ref{Dis} and \ref{Ove}. As seen in Fig.\ \ref{Dis}, the distributed control scheme works well even when there are 5000 houses in the area. The main reason is that multiple power control loops are conducted simultaneously at all clusters in the system, which equivalently leads to the reduction of the total time required for controlling the whole system. However, the performance of the distributed control scheme varies with the cluster size. As seen in Fig.\ \ref{Ove}, the distributed control with 9 clusters (approx. 555 houses/ cluster) can perfectly avoid the consumed power overload while that with 4 clusters (approx. 1250 houses/cluster) has a slight power excess over the target limit of 1080W. It is  obvious that the larger cluster size has the worse performance, or equivalently leads to inferior control perfomance of the overall system. On the other hand, a distributed control system with many tiny-size clusters is not reasonable due to the following two reasons. Firstly, each sub-controller of the tiny-size cluster has only very few knowledge of the overall system, which makes the control of each cluster become diversified, thus the cooperative control of all clusters becomes loose. Secondly,  data traffic burden on the backbone network will increase with the number of clusters, which is not desirable because of  limited capacity of the backbone networks  for Smart Grid. Based on the above discussion, we could confirm the validity of proposed distributed control system and recommend approximately 500 houses per cluster in this design.

Futhermore, Fig.\ \ref{powerdistribution} shows the distritribution of instaneous power consumption of the overall 5000 houses during the simulation time. It can be observed that the number of houses consuming more than the average 1200W in the centralized control case is higher than that in the distributed case.  In the centralized control case, there is higher probability that some home-gateways could not control properly due to limited coverage and longer delay. In the distributed control case, most of the home-gateways succeed to control the load demand to reduce their power usage of low priority  appliances to avoid overload. As the result, the power consumption during peak hours converges to some specific values, e.g. 300W, 1130W etc., which coressponds to the consumed power of appliances with high priority.

\section{Green Building Test-Bed System}

\begin{figure} [h]
\centering
\includegraphics[width=8.5cm,height=10cm]{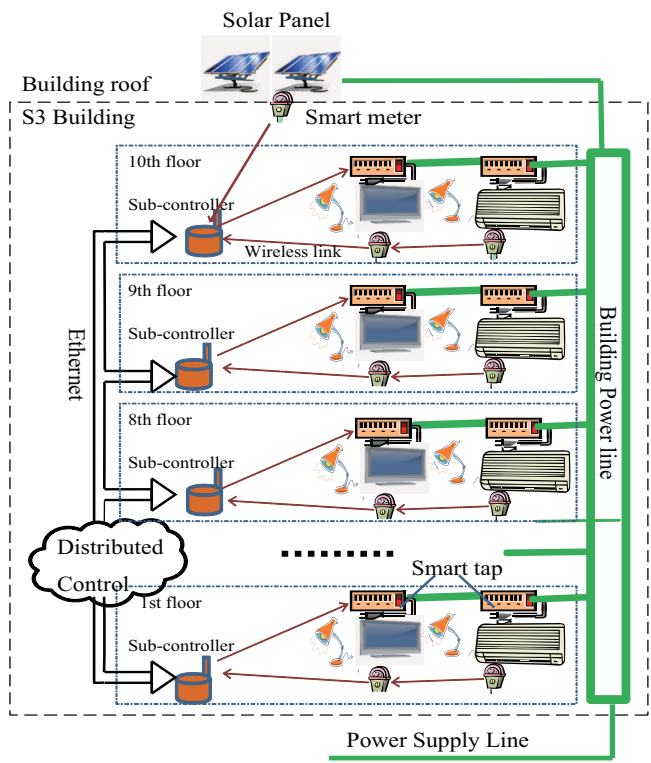} 
\caption{Green building test-bed system.}
\label{Testbed} 
\end{figure}

\begin{figure} [h]
\centering
\includegraphics[width=8.5cm]{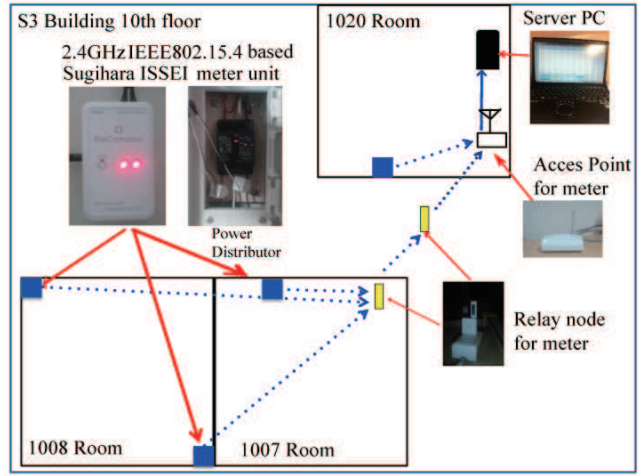} 
\caption{Smart meter network for green building test-bed.}
\label{SmaNet} 
\end{figure}

\begin{figure} [h]
\centering
\includegraphics[width=8.5cm]{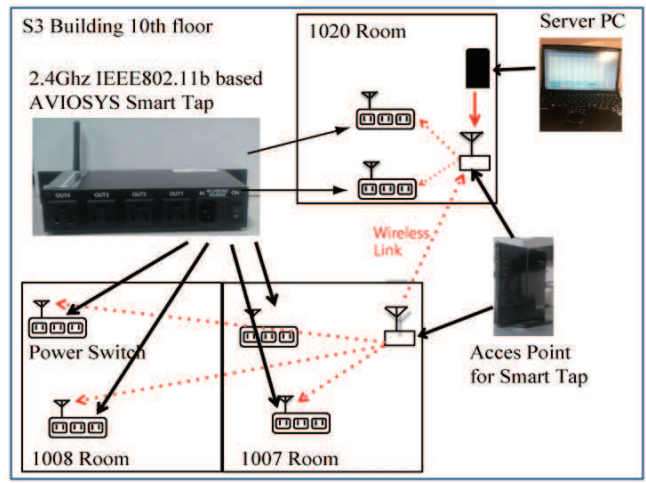} 
\caption{Actuator network for green building test-bed.}
\label{ActNet} 
\end{figure}

 A building for the Department of Electrical $\&$ Electronic Engineering in Tokyo Institute of Technology is used for the test-bed implementation. This  small-scale test-bed is sufficient to check the feasibility of the proposed algorithm since it is scalable against the number of nodes in the network as explained in Sec. 2. Figure\ \ref{Testbed} shows the whole system structure, in which each laboratory-scale power system consists of a sub-controller installed in a PC server in charge of laboratory power management, a smart meter network,  and a load power control network. Figure\ \ref{SmaNet} shows the installed smart meter network based on IEEE 802.15.4 standard in our laboratory at 10th floor. The smart meter is located at a circuit breaker box in each room to measure the consumed power every 5s with a resolution of 10W. The smart meter sends the measured data to the server via wireless link of IEEE 802.15.4 using multi-hop topology. Figure \ref{ActNet} shows the installed load power control network using IEEE 802.11b WLAN in our laboratory. All electric appliances in each room are connected to power switches which are remotely controllable via IEEE 802.11b standard.

The smart meter network measures the power consumption of all electric appliances in each laboratory. If the total load power becomes larger than a power threshold, each sub-controller locally turns off the power switch connected to each appliance via the load power control network. In the power switch, there are four ports with different priorities. In the case of power saving, the server starts to turn off the switch connecting to the appliance with lowest priority. The power threshold in the central controller is determined based on the average power consumption data in previous year, power saving ratio, and current supply power from solar power generator.

\begin{figure} [h]
\centering
\includegraphics[width=8.5cm]{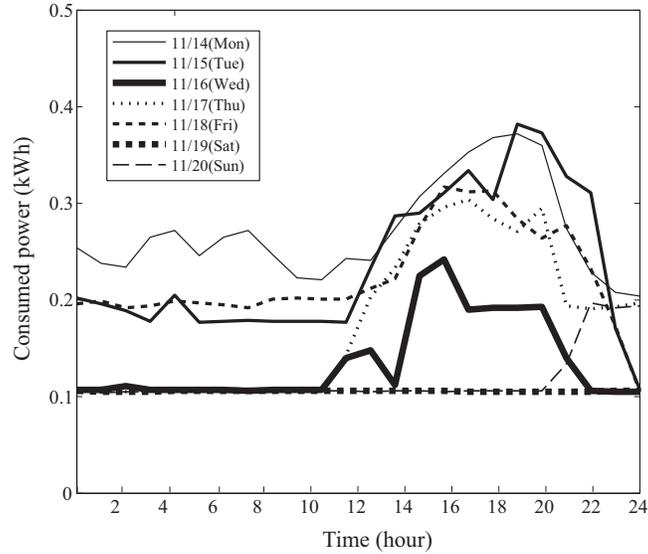} 
\caption{Power consumed in student room  during a week.}
\label{Fweekdata} 
\end{figure}

 \begin{figure} [h]
\centering
\includegraphics[width=8.7cm]{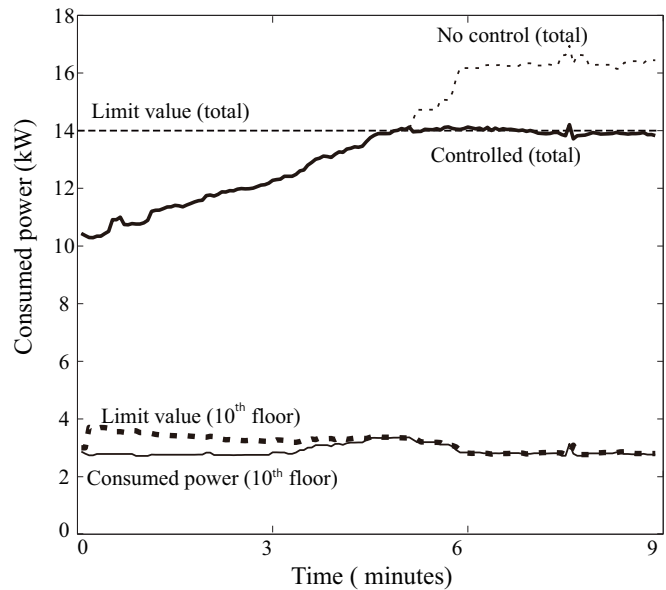} 
\caption{Results on performance of testbed system.}
\label{Meaadd} 
\end{figure}

Figure\ \ref{Fweekdata} shows an example of the variation of consumption power in the room 1020 over a week. Nominally, six students are in the room. The figure reveals that the students come around noon and leave the room around 22 o\textquoteright clock. There is no one in the room on Sunday, and the power consumption of 100W on this day corresponds to standby power of devices.

 Finally, Fig.\ \ref{Meaadd} shows the performance of power saving with and without the proposed power control applying to the four laboratories in the buiding. In this setup, a total power threshold of 14kW is artificially employed during the peak hour around 14 o\textquoteright clock. The distributed control algorithm is implemented in all four laboratory sub-controllers to fulfill the total power threshold. The total power consumptions of four laboratories with and without control, the power consumption and the assigned power limit in our laboratory are illustrated in the Fig.\ \ref{Meaadd}. The total power consumption increases from normal power usage of 10kW to 14kW in approximately five minutes. After that the control is applied to suppress the excess consumption power. At the same time, the  power consumption limit corresponding to our laboratory is calculated by the proposed algorithm and reassigned every 10s. As seen in the figure, the controller of our laboratory starts to reduce the consumption power as soon as the total consumption power exceeds to the threshold value, which contributes to avoiding  power overload in peak hours. In conclusion, these results  indicate the effectiveness of the developed control system in real applications.

\section{Conclusion}
This paper took the first step toward establishing hierarchical distributed architecture for large-scale power control system to realize demand and response during peak hours. In the proposed architecture, there are many sub-controllers in charge of managing the power consumption in their own clusters of approximately 500 houses. The sub-controllers are subject to local power consumption limits assigned from upper layer, which forms a local control loops small enough to guarantee stable control. In addition, we proposed a distributed control algorithm where sub-controllers at higher layers determine appropriate local power consumption limits, which contributes to realizing the global objective of power reduction during peak hours. We showed that the proposed control network is scalable regardless of the size of the power system through numerical simulations with realistic parameters. Furthermore, a building-scale test-bed for power control system was implemented to show the effectiveness of the proposed scheme contributing to daily life power saving instead of high-cost planned blackouts. Our future work will focus on the impact of integrating a large number of electric vehicles, distributed batteries and other renewable energy sources to the system for improvement of demand response performance.



\end{document}